\documentclass{jltp}

\usepackage{graphicx} 

\newcommand{\figwidth}{12.5cm}
\newcommand{\bm}[1]{\mbox{\boldmath $#1$}}

\def\vn{{\bm v}^n}
\def\vs{{\bm v}^s}
\def\u{\bm u}
\def\bnabla{\bm{\nabla}}
\def\bcdot{\bm{\cdot}}
\def\nus{\nu^s}

\def\etal{\mbox{\it et al}\,}

\def\bOmega{\bm \Omega}
\def\blambda{\bm \lambda}
\def\i{\mbox{\rm i$\,\!$}}
\def\etc{\mbox{\it etc}}
\renewcommand{\Im}{{\cal I}m}
\renewcommand{\Re}{{\cal R}e}

\title{Torsional Oscillations of a Rotating Column of $^3$He-B}

\author{Karen Henderson and Carlo Barenghi$^*$}

\address{School of Mathematical Sciences, CEMS, University of the 
  West of England, \\ Bristol, BS16 1QY, UK\\
$^*$School of Mathematics and Statistics, University of Newcastle,
\\  Newcastle upon Tyne, NE1 7RU, UK}

\runninghead{K. Henderson and C. Barenghi}{Torsional Oscillations 
of a Rotating Column of $^3$He-B}

\begin{document}

\maketitle

\begin{abstract}
We have analysed the axisymmetric and
non-axisymmetric modes of a continuum of vortices in a rotating
superfluid.  We have investigated how changing the temperature affects
the growth rate of the disturbances.  We find that, 
in the long axial wavelength limit 
the condition $q=\alpha/(1-\alpha')=1$, where $\alpha$ and $\alpha'$ 
are temperature-dependent 
mutual friction parameters, is the crossover between damped and 
propagating Kelvin waves.
Thus 
at temperatures for which $q>1$, perturbations on the vortices are unlikely
to cause vortex reconnections and turbulence.
These results are in agreement with the recent discovery of 
Finne \etal\cite{fin.etal:03} of an intrinsic
condition for the onset of quantum turbulence in $^3$He-B.

\end{abstract}

\section{MODEL}

When superfluid helium is rotated, an array of quantised vortex lines 
appear which are aligned parallel to the axis of rotation.  This array 
exhibits oscillation modes which were first predicted and
observed by Hall\cite{hall:60}.  
The modes are similar to classical inertial waves 
in the long wavelength limits and are related to the waves of isolated vortex
lines in the short wavelength limit.  
These vortex waves are a very important phenomena in the understanding
of quantised vortex lines and for a review of work in this area see
Donnelly\cite{don:91}.

The aim of this paper is to extend previous work\cite{hen.bar.euro:04} 
to consider the torsional oscillations of a rotating column of superfluid
at a range of temperatures.
We consider $^3$He-B rotating with angular velocity $\bm{\Omega}=\Omega {\bf{e}}_z$ in a cylinder of radius $r=a$.  The basic state will
consist of
a uniform vortex lattice with a large density of vortices aligned along
the direction of rotation. 
In this context the behaviour of the
superfluid is described by the Hall-Vinen equations\cite{hall.vin:56}. Thus, rather
than individual vortex lines, we consider a continuum of vortex lines.
The advantage of this model is that it allows us to explore effects 
which the theory of a single vortex filament\cite{bar.don.vin:85} 
cannot describe, notably 
the presence of boundaries and the degrees of freedom represented by the 
coherent oscillatory motion of many vortices. 

The equations of motion of the superfluid in a coordinate system rotating 
with angular velocity $\bm{\Omega}=\Omega {\bf{e}}_z$ may be written as
\begin{eqnarray}
 \frac{\partial \vs}{\partial t}
\!\!&\!\!+\!\!&\!\! (\bm{\vs \cdot\nabla }) \vs
=  \bnabla \Psi +
2\vs\times\bOmega +\alpha\widehat{\blambda}\times[\blambda \times(\vs-\vn)]
\nonumber \\[-1.5ex]
&& \label{eq:vs}\\[-1.5ex]
\nonumber
&&+\alpha'\blambda \times(\vs-\vn)- \alpha\nus\widehat{\blambda}\times
(\blambda \bcdot \bnabla)\widehat{\blambda}
+\nus(1-\alpha')(\blambda \bcdot \bnabla)\widehat{\blambda}\nonumber 
\end{eqnarray}
where $\vs$ and $\vn$ are the superfluid and normal fluid velocities in 
the rotating frame, $\blambda=\bnabla \!\times \!\vs + 2\bOmega$, 
$\widehat{\blambda}=\blambda/|\blambda|$ is the unit vector in the
direction of $\blambda$ and $\Psi$ is a collection of scalar 
terms.  Given the high viscosity of $^3$He-B, we assume that the normal fluid
is in solid body rotation around the $z$-axis, thus in the rotating frame
$\vn=0$.  Equation~(\ref{eq:vs})
must be solved under the condition that $\bnabla \!\bcdot \!\, \vs=0$.
The quantity 
$\nus=(\Gamma / 4\pi) \log (b_{_0}/a_{_0})$ is the vortex tension
parameter, $\Gamma$ is the quantum of circulation, $a_{_0}$ is the
vortex core radius and $b_{_0}=(|\blambda|/\Gamma)^{-1/2}$ is the
average distance between vortices. 
The unperturbed vortex lattice
corresponds to the basic state $\vs_0=\vn_0=0$, $\nabla\Psi_0=0$
for which $\blambda=2\Omega {\bf{e}}_z$, working in cylindrical
coordinates $(r,\phi,z)$.

We perturb the basic state by letting $\vs=\u=(u_r,u_\phi,u_z)$, 
$\Psi=\Psi_0+\psi$,
where $|\u|\ll 1$ and $|\psi|\ll 1$  and
linearise the resulting equations of motion. We 
assume normal modes for the perturbations of the form
 $(\u,\psi)=(\widehat{\bm{u}}(r),\widehat{\psi}(r))
  \exp(\i\sigma t+\i m\phi + \i kz)$
where $m$ and $k$ are respectively
the azimuthal and axial wavenumbers and $\sigma$ is the growth rate.
The aim of our calculation is to determine the real and imaginary parts
of $\sigma$, namely $\Re(\sigma)$ and $\Im(\sigma)$.

The solution for $\widehat{u}_z$  which is regular as $r\to 0$ is
the Bessel function of the first kind of order $m$, $J_m(\beta r)$,
where
\begin{equation}
\beta^2=\frac{-k^2[(\i\sigma+\alpha \eta)^2 +(1-\alpha')^2 \eta^2]}
{[(\i\sigma+\alpha \eta)(\i\sigma+k^2 \nus \alpha)
+(1-\alpha')^2\nus k^2 \eta]}
\label{eq:beta}
\end{equation}
and $\eta=2\Omega+\nus k^2$.
\noindent
To determine $\beta$ we enforce the boundary condition $u_r=0$
on the wall of the container $r=a$, which yields the secular equation
\begin{equation}
\frac{(ka)^2}{(\beta a)}\frac{J'_m(\beta a)}{J_m(\beta a)} 
[(\i\sigma +\alpha \eta)^2 +(1-\alpha')^2\eta^2]
+2m\Omega \sigma (1-\alpha')=0,
\label{eq:eigen}
\end{equation}
\noindent
where the prime denotes the derivative of the Bessel function with respect to
its argument.

\section{RESULTS}

Equations~(\ref{eq:beta},\ref{eq:eigen}) simplify in the axisymmetric
case, $m=0$. From (\ref{eq:eigen}) we find that $\beta$ must satisfy
$J'_0(\beta a)=J_1(\beta a)=0$.  Taking $\xi_j$ to be the $j^{th}$ zero of 
$J_1(\xi)=J'_0(\xi)$
(that is $\xi_0=0$, $\xi_1=3.83171,$ $\xi_2=7.01559$ \etc),
Equation~(\ref{eq:eigen}) becomes
\begin{equation}
\frac{\xi^2_j}{(ka)^2}=
-\frac{[(\i\sigma +\alpha \eta)^2+(1-\alpha')^2\eta^2]}
{[(\i\sigma+\alpha \eta)(\i\sigma+\alpha \nus k^2)
+(1-\alpha')^2\nus \eta k^2]}.
\label{eq:axi1}
\end{equation}
This quadratic equation may be solved 
analytically\cite{gla.etal:74,hen.bar.euro:04} 
and has an infinite number of solutions, according to the
value of $\xi_j$ considered.
All the solutions are such that $\Im(\sigma)$ is non-negative,
so the system is always stable to infinitesimal disturbances.  
The least 
stable mode is the one for which $\Im(\sigma)$ is minimum and is
given by
\begin{equation}
\sigma =\i\alpha(\Omega +\nus k^2)
\pm \sqrt{\left\{(1-\alpha')^2\nus k^2(2\Omega+\nus k^2)
-\alpha^2\Omega^2\right\}}
\label{eq:xinf}
\end{equation}
which corresponds to $\xi_j \rightarrow \infty$.

\begin{figure}
\begin{center}
\includegraphics[width=\figwidth]{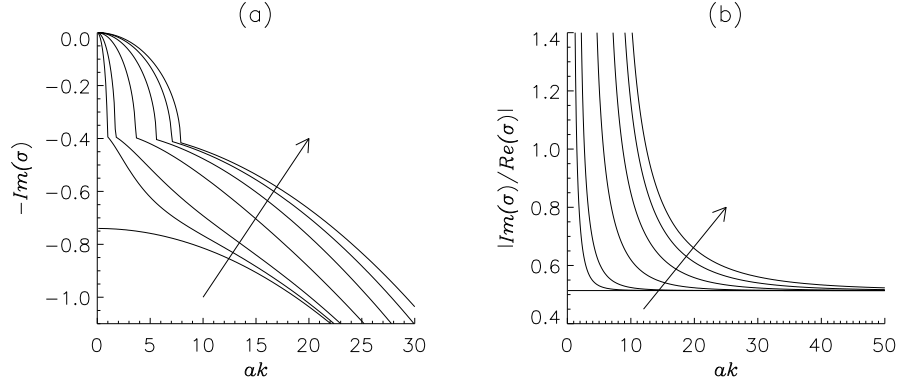}
\caption{Plots of (a) the growth rate, $-\Im(\sigma)$ 
(b) $|\Im(\sigma)/\Re(\sigma)|$ of the axisymmetric mode $(m=0)$
against $ak$ for various values of $\xi_j$ at $T/T_c=0.5$.
The arrows show the direction of increasing $\xi_j$, where 
$j=0,1,2,5,10,20,\infty$.}
\label{fig:m0}
\end{center}
\end{figure}

\begin{figure}
\begin{center}
\includegraphics[width=\figwidth]{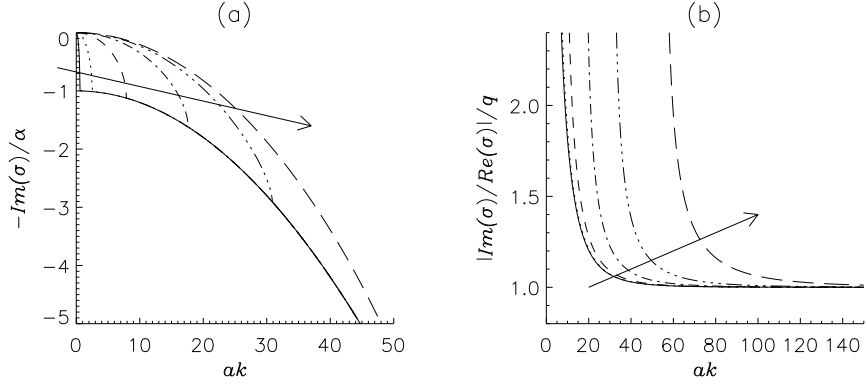}
\caption{Plots of (a) $-\Im(\sigma)/\alpha$ 
(b) $|\Im(\sigma)/\Re(\sigma)|/q$ of the least stable axisymmetric mode
against $ak$ for various temperatures.
The arrows show the direction of increasing $T$, where 
$T/T_c=0.3,0.4,0.5,0.6,0.7,0.8$.}
\label{fig:m0T}
\end{center}
\end{figure}

Plots of the growth rate, $-\Im(\sigma)$ and the ratio of the decay 
rate of the wave to its angular frequency, $|\Im(\sigma)/\Re(\sigma)|$ can 
be found in Fig.~\ref{fig:m0}, where the mutual friction parameters 
correspond to $T=0.5T_c$ for which $\alpha=0.38$ and $\alpha'=0.28$, 
where $T_c$ is the critical temperature
and we have taken $\nus/a^2=0.002$ and $\Omega=1$.  
The kinks in the curves of Fig.~\ref{fig:m0}(a) for $j>0$ correspond
to $\sigma$ switching from being purely imaginary (for which the
decay of the mode is monotonic with time) to complex (for which the
decay of the mode is oscillatory).
For the least stable mode, the crossover occurs at $k=k_\ast$ where
\begin{equation}
 k_\ast^2=\frac{\Omega}{\nus}(-1+\sqrt{1+q^2}) 
\end{equation}
where $q=\alpha/(1-\alpha')$.  The parameter $q$ increases rapidly
with temperature\cite{bev.etal:97} and
therefore $k_\ast$ also increases rapidly with temperature.
This effect is illustrated in Fig.\ref{fig:m0T}(a) in which,
for clarity of scale, we plot $-\Im(\sigma)/\alpha$ against $(ak)$
of the least stable axisymmetric mode at various temperatures.
The arrow shows the direction of increasing temperature.
For $k>k_\ast$, the expression in the square root of (\ref{eq:xinf})
is positive so the normalised growth rate is independent of
temperature and is given by $\Im(\sigma)/\alpha=\Omega+\nus k^2$.
The mutual friction parameter $\alpha$ increases rapidly with 
temperature so the axisymmetric modes will decay much faster 
at higher temperatures than at lower temperatures.  
In Fig.\ref{fig:m0}(b) we plot the ratio of the decay rate of the 
wave to its angular frequency, $|\Im(\sigma)/\Re(\sigma)|$
against $(ak)$  We see that for all the modes 
$|\Im(\sigma)/\Re(\sigma)|\rightarrow q\approx 0.51$ at large $(ak)$.
In Fig.\ref{fig:m0T}(b) we plot $|\Im(\sigma)/\Re(\sigma)|/q$
against $(ak)$ of the least stable axisymmetric mode at various
temperatures.  Again we find $|\Im(\sigma)/\Re(\sigma)|\rightarrow q$
at large $(ak)$ for all temperatures and that this limiting value is 
obtained at larger values of $(ak)$ as the temperature is increased.

\begin{figure}
\begin{center}
\includegraphics[width=\figwidth]{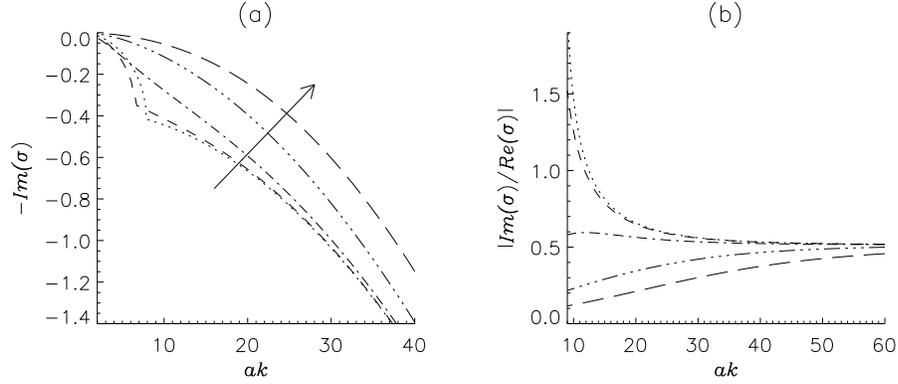}
\caption{Plots of (a) the growth rate, $-\Im(\sigma)$ 
(b) $|\Im(\sigma)/\Re(\sigma)|$ against $ak$ for the $m=0,2,4,7,10$
mode. The arrow shows the direction of increasing $m$.} 
\label{fig:allm}
\end{center}
\end{figure}

In order to consider the non-axisymmetric modes we must solve the 
coupled equations (\ref{eq:beta},\ref{eq:eigen}) and further details 
of the solution method can be found in Henderson \& 
Barenghi~\cite{hen.bar.euro:04}.
In Fig.~\ref{fig:allm} we plot $-\Im(\sigma)$ and 
$|\Im(\sigma)/\Re(\sigma)|$ against $ak$ of the least stable computed modes
for $m=0,2,4,7,10$ using the same parameters as for Fig.~\ref{fig:m0}. 
The arrow shows the direction of increasing $m$. We do not plot the $m=1$
for clarity; it suffices to say that the least stable $m=1$ mode is bound
by the least stable $m=0$ mode.  

It can be seen that the modes become less 
stable as $m$ is increased which is similar to what we found at lower 
temperatures\cite{hen.bar.euro:04}.
We can also see that at large $(ak)$, all the modes are such that
$|\Im(\sigma)/\Re(\sigma)|\rightarrow q \approx 0.51$ and that this limit
is approached from below for larger values of $m$.  What this means is that 
oscillatory decay is more pronounced for modes with larger azimuthal 
wavenumbers, particularly at small to moderate values of the axial wavenumber.
As for the azimuthal mode, we find that the decay rates of the azimuthal
modes, $\Im(\sigma)$ are
proportional to $\alpha$, so at higher temperatures, for which $\alpha$ is
small, the mode will decay more rapidly in time.

\section{APPLICATION}

Our work has applications to the recent discovery by 
Finne \etal\cite{fin.etal:03} of an intrinsic condition for the onset of 
quantum turbulence in $^3$He-B, namely $q>1.3$.
Finne \etal $\,$ interpreted their result in terms of Kelvin waves - helical 
perturbations of the position of a vortex core away from the unperturbed 
straight shape.
In the case of an isolated vortex line, it was predicted \cite{bar.don.vin:85}
that $q=1$ is the crossover between Kelvin waves which propagate ($q<1$)
and Kelvin waves which are damped ($q>1$). If $q<1$ the propagating
Kelvin waves grow in amplitude exponentially with time, 
driven by the local difference between the normal fluid velocity and the 
velocity of the vortex line. 

If this condition is valid for
a large number of vortices (which have more degrees of freedom that a
single vortex line), then one expects that, when
the waves' amplitude becomes of the order of the average intervortex
spacing, the vortices reconnect which each other, and form a
turbulent tangle.

Our analysis shows that, 
for both axisymmetric and non-axisymmetric perturbations of
a vortex lattice, 
the ratio of imaginary and real part of the
complex growth rate is approximately equal to the parameter $q$ identified
by Finne \etal $\,$ in the large axial wavenumber limit,
for all temperatures.
The minimum permitted value of $k$ will be governed by the height, $h$ of
the apparatus by the relation $k_{\rm min}=2\pi/h$, so 
provided that the aspect ratio is small enough ($h/a\ll 1$) we find
agreement with the argument of Finne \etal.
Therefore if $q>1$, perturbations
shrink in amplitude before they can rotate a full cycle (overdamping);
this reduces the volume of space which is swept by the vortex lines
in their motion, which reduces the probability of making 
reconnections with neighbouring vortices, and without reconnections
there is no turbulence. Vice versa,
if $q<1$ the perturbations propagate, which favours vortex reconnections,
hence turbulence. 

Finally, the exact dispersion relation which we have found
(\ref{eq:beta},\ref{eq:eigen}) can be used to study with
precision the torsional oscillations of on a rotating vortex lattice for
any height and radius of experimental apparatus, which is a topic of
current experimental interest\cite{Golov}.





\end{document}